\documentclass[11pt]{article}
\usepackage{moriond,epsfig}

\def\PLB{{\em Phys. Lett.}  B}
\def\PRL{{\em Phys. Rev. Lett.}}
\def\PRD{{\em Phys. Rev.} D}

\def\be{\begin{equation}}
\def\ee{\end{equation}}
\def\bea{\begin{eqnarray}}
\def\eea{\end{eqnarray}}

\begin{document}
\title{NEW PHYSICS SEARCH AT LHCB}

\author{ A.HICHEUR\footnote{Adlene.Hicheur@cern.ch}\\
{\it On behalf of the LHCb collaboration}}

\address{\'Ecole Polytechnique F\'ed\'erale de Lausanne - LPHE,\\
Lausanne 1015, Switzerland.} \maketitle

\abstract{ Although direct detection of new particles will be the
main focus of the LHC, indirect New Physics searches are expected
to provide useful complementary information. In particular,
precision measurements of rare processes occurring in flavour
physics are of utmost importance in constraining the structure of
the New physics low energy effective Lagrangian. In this paper,
few key LHCb studies, including $B_s-\bar{B}_s$ mixing and rare
decays through the quark level $b\to s$ loop transition, are
presented to illustrate New Physics effects at low energy.}

\section{Introduction}
Weak decays of hadrons are generically described by a low energy
effective hamiltonian expressed as an expansion of local operators
$O_i$ \cite{Buchalla_weak}:
\begin{equation}
H_{eff}=\sum_i C_i O_i
\end{equation}
The Wilson coefficients $C_i$ include the short distance effects
and are computed perturbatively at the electroweak scale and then
derived at the $\sim m_b$ scale through the renormalization group
equations. The matrix elements of the $O_i$ operators represent
the long range effects related to hadronization and are derived
non-perturbatively, using various techniques (QCD sum rules,
Lattice, etc...). Note that the $O_i$ also mix under
renormalization, the consequence being that a given $C_i$
coefficient may receive contributions from other $C_j$
coefficients: in this case, we talk about effective
coefficients, $C_i^{eff}$ associated to $O_i$.\\
In this framework, the intervention of virtual new heavy particles
in loop dominated processes will affect the $C_i$ coefficients. We
are therefore in search for any observable sensitive to these
coefficients and for which the theoretical uncertainties are
relatively small.
\section{$B_s$ mixing}
\begin{figure}[!h]
\centering \psfig{figure=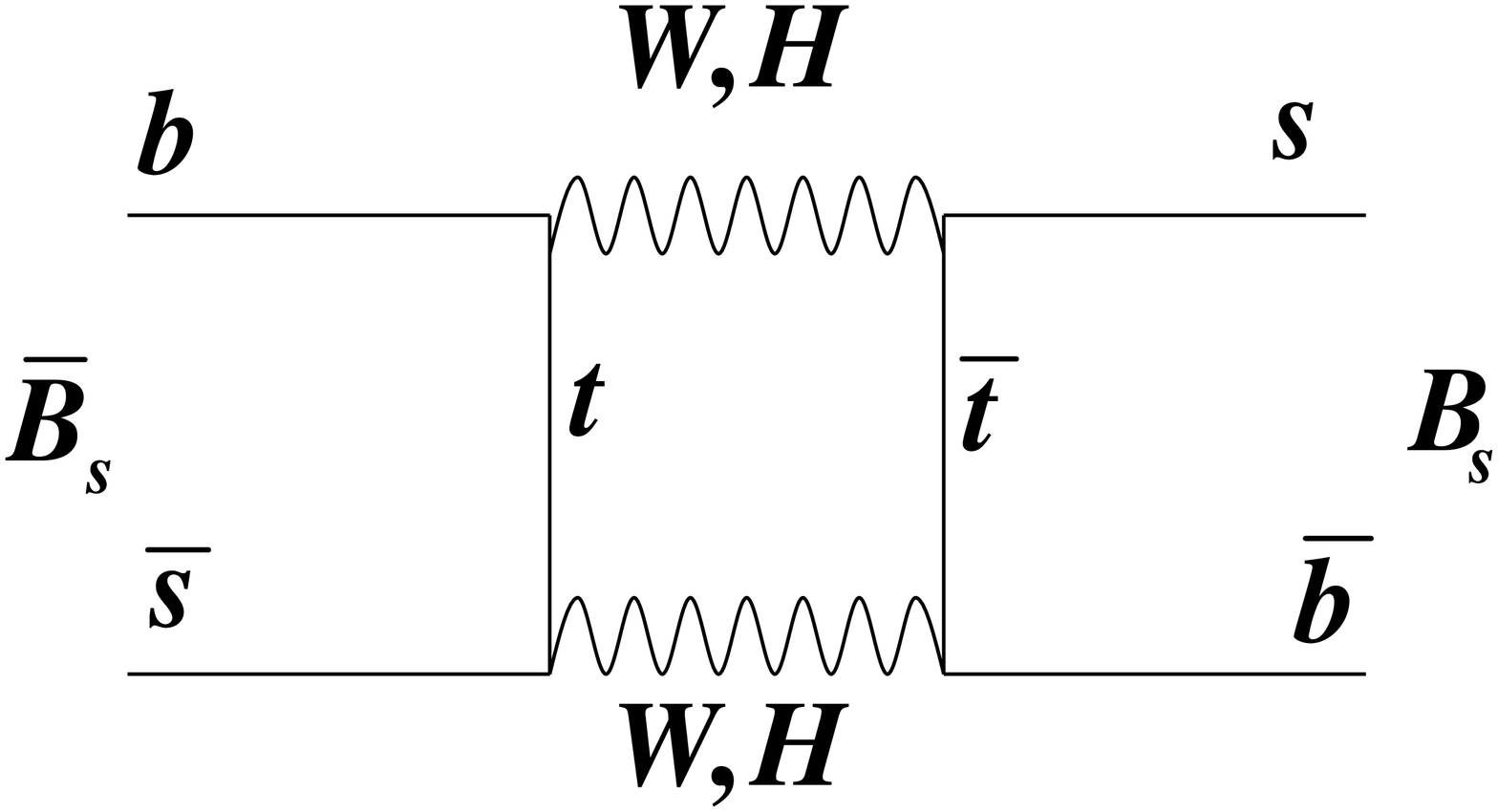,scale=0.2}
\psfig{figure=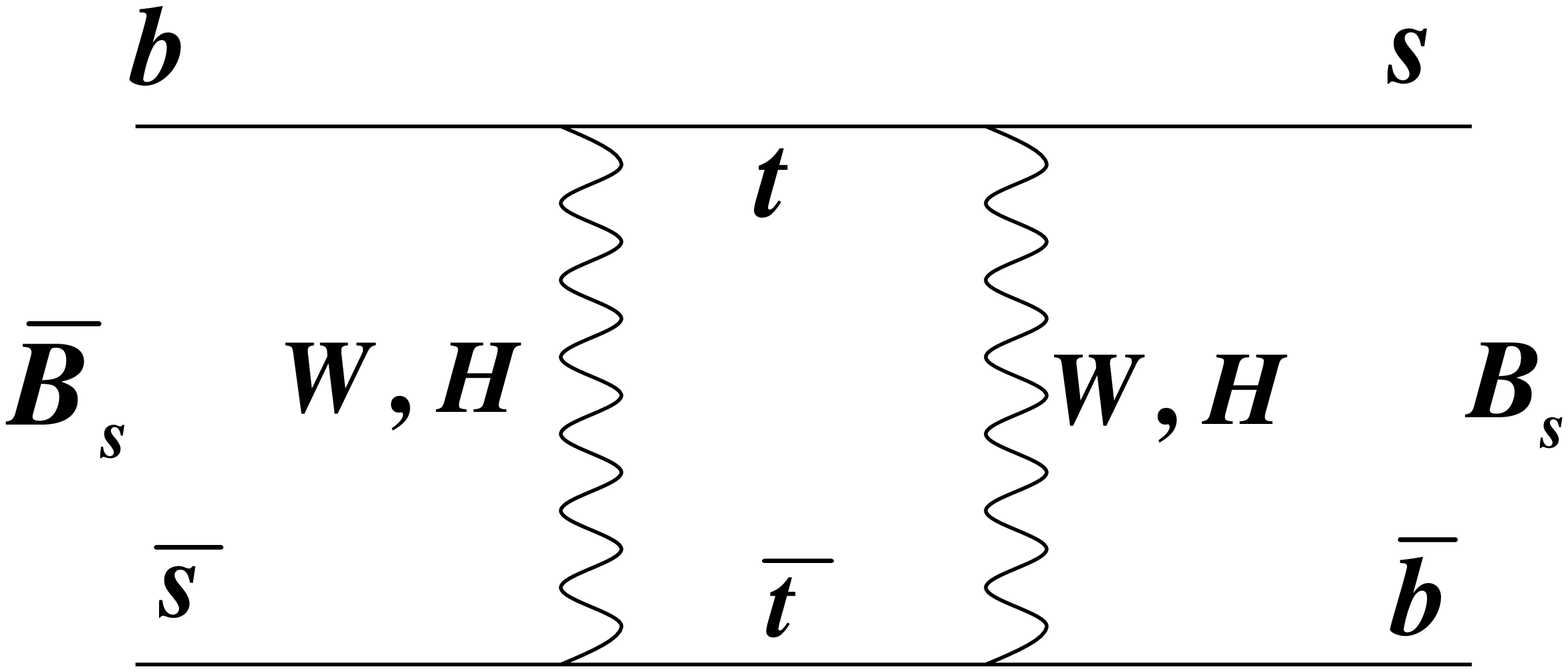,scale=0.2}
 \caption{Box
diagrams for $B_s$ mixing. In Standard Model, the loop is mediated
by a $W$ boson. For illustration, the case where a charged Higgs
is involved is depicted.\label{fig:bsmixingdiag}}
\end{figure}

The $B_s-\bar{B}_s$ meson oscillation is described by the
$\Delta~B=2$ box diagrams shown in figure \ref{fig:bsmixingdiag}.

In the Standard model, the box diagrams carry the weak phase
$(V_{tb}V_{ts}^*)^2$.
 The best way to probe new contributions in
the box is to compare $B_s$ and $\bar{B}_s$ decays to common $CP$
final states as a function of the $B_s$ proper time. Similarly to
the $B_d$ case, the preferred final states $f_{CP}$ are the ones
induced by the $b\to c\bar{c}s$ quark tree transition, leading to
the golden mode $B_s\to J/\Psi \phi$ and other modes less favored
experimentally such as $B_s\to J/\Psi \eta^{(\prime)}$,
$B_s\to\eta_c\phi$ and $B_s\to D_s^+D_s^-$. All these modes carry
the weak phase $V_{cb}V_{cs}^*$.\\
The amplitude of the mixing-induced asymmetry, $A_{CP}(t) =
\frac{\Gamma(\bar{B}_s(t)\to f_{CP})-\Gamma(B_s(t)\to
f_{CP})}{\Gamma(\bar{B}_s(t)\to f_{CP})+\Gamma(B_s(t)\to f_{CP})}$
is proportional to $sin(2\beta_s)$, where
$2\beta_s=2arg(-\frac{V_{ts}V_{tb}^*}{V_{cs}V_{cb}^*})=-0.04$ to a
precision of 5\% in the Standard Model. This small value implies
that the measurement of a sizeable amplitude will point directly
to the intervention of New Physics.\\
Table \ref{tab:Betas_sensitivity} shows results of sensitivity
studies performed for various decay channels with the LHCb
simulated events. The numbers have been obtained for a statistics
corresponding to one year of data taking at nominal luminosity
$L=2\times 10^{34} cm^{-2}.s^{-1}$.

\begin{table} [!h]
{\begin{center}

\begin{tabular}{|p{5.2cm}|p{4cm}|p{4cm}|}
\hline \hspace{1.5cm} \textbf{Sample} &  Expected yield/$2~fb^{-1}$ & \hspace{1.5cm} $\sigma(2\beta_s)$ \\
\hline $B_s\to
J/\Psi(\mu^+\mu^-) \eta(\gamma\gamma)$ \cite{fernandez_thesis} & \hspace{1.5cm}8.5k & \hspace{1.5cm}0.109\\
\hline $B_s\to
J/\Psi(\mu^+\mu^-) \eta(\pi\pi\pi^0)$ \cite{fernandez_thesis}& \hspace{1.5cm}3k & \hspace{1.5cm}0.142\\
\hline $B_s\to
J/\Psi(\mu^+\mu^-) \eta^{\prime}(\pi\pi\eta)$ \cite{jpsietap_etapipi}& \hspace{1.5cm}2.2k & \hspace{1.5cm}0.154\\
\hline $B_s\to
J/\Psi(\mu^+\mu^-) \eta^{\prime}(\rho\gamma)$ \cite{jpsietap_rhogam}& \hspace{1.5cm}4.2k & \hspace{1.5cm}0.080\\
\hline $B_s\to
\eta_c(4h) \phi(K^+K^-)$ \cite{fernandez_thesis} & \hspace{1.5cm}3k & \hspace{1.5cm}0.108\\
\hline $B_s\to
D_s^+ D_s^-$ \cite{fernandez_thesis}& \hspace{1.5cm}4k & \hspace{1.5cm}0.133\\
\hline\hline All pure $CP$ eigenstates & \hspace{1.5cm}-& \hspace{1.5cm}0.046\\
 \hline \hline
 $B_s\to J/\Psi(\mu^+\mu^-)\phi(K^+K^-)$ \cite{fernandez_thesis,threeangles_phis}& \hspace{1.5cm}130k & \hspace{1.5cm}0.023\\
\hline \hline All modes & \hspace{1.5cm}-& \hspace{1.5cm}0.021\\
\hline
  \end{tabular}
\caption{\label{tab:Betas_sensitivity} Expected yields and
sensitivities on $2\beta_s$. Reconstructed submodes are indicated
between brackets. For $\eta_c$ reconstruction, $4h$ means a
combination of four charged kaons or pions.}
  \end{center}
}
\end{table}

For a short term scenario with an integrated luminosity of
$0.1~fb^{-1}$ the sensitivity to $2\beta_s$ is still better than
$0.1$, which means that a deviation of $\sim 0.3$ would be
detected with very early data.\\
New physics contribution to the mixing is usually parameterized in
a model-independent way \cite{Buchalla_newphys}:
\begin{equation}
\frac{<B_s|H_{eff}^{total}|\bar{B}_s>}{<B_s|H_{eff}^{SM}|\bar{B}_s>}=C_{B_s}e^{2i\phi_{B_s}}
\end{equation}
It follows from this parameterization that $sin(2\beta_s)$ becomes
$sin(2(\beta_s-\phi_{B_s}))$. Fits for $C_{B_s}$ and $\phi_{B_s}$
parameters have been performed using available experimental data
\cite{UTfit_phis_first}. The recent Tevatron results on the mixing
phase and width difference $\Delta\Gamma_s$ \cite{D0Collab_phis},
$-1.20 < 2\beta_s < 0.06$ and $0.06< \Delta\Gamma_s <0.30~ps^{-1}$
at 90\% confidence level, triggered a statistical analysis
\cite{UTfit_phis_last} which allowed to constrain the
$C_{B_s}-\phi_{B_s}$ parameters space, suggesting a hint for
beyond SM contributions. No doubt that the coming improvements in
Tevatron results and above all, the first LHCb results, will
definitely clarify the picture and help us quantify more
accurately the size of any New Physics contribution.

\section{Radiative $b\to s\gamma$}
$b\to s\gamma$ is one of the benchmark New Physics probe in $b$
physics. It is mediated by the electromagnetic dipole operator,
$O_7 = \bar{s}\sigma^{\mu\nu}(m_bR+m_sL)bF_{\mu\nu}$, where
$R=1+\gamma^5,~L=1-\gamma^5$. The amplitude is therefore driven by
the effective Wilson coefficient $C_7^{eff}$.

Given that $m_s<<m_b$, one photon polarization is suppressed by
$m_s/m_b$: the photon is mostly right-handed in $\bar{b}$ decays
and left-handed in $b$ decays. However, enhancement of the
suppressed polarization could come from New Physics contributions.
Furthermore, it can be shown that the mixing-induced asymmetry in
$B^0\to V^0\gamma$ decays has the
same suppression factor as the photon polarity. \\
Several radiative decays studies have been performed in LHCb,
among them: $B_d\to K^{*0}(K^+\pi^-)\gamma$, and
$B_s\to\phi(K^+K^-)\gamma$ \cite{KstarPhiGam_Lhcb}. These modes
have been jointly analyzed and common selection cuts have been
applied, when possible. For the photon, a ECAL cluster not
associated to a track is required, along with a transverse energy
cut to suppress $\pi^0$ background. To reconstruct $K^{*0}$ and
$\phi$, impact parameters and particle ID cuts are applied to
pions and kaons, as well as vertex quality requirements for $K\pi$
and $KK$. $B$ flight is used to reject prompt background from
primary production vertex.

The studies have shown that yields of 68k and 11k signal events
are expected with $2~fb^{-1}$ for $B_d\to K^{*0}\gamma$ and
$B_s\to\phi\gamma$, respectively. With this statistics, a 1\%
sensitivity is expected for the $CP$ asymmetry. A dedicated photon
polarization study was performed for the $B_s\to\phi\gamma$
channel \cite{PhiGamPol_Lhcb} and has shown a sensitivity better
than 0.2 for the suppressed polarization fraction.
\section{Electroweak $b\to s l^+l^-$}
\begin{figure}[!h]
\centering \psfig{figure=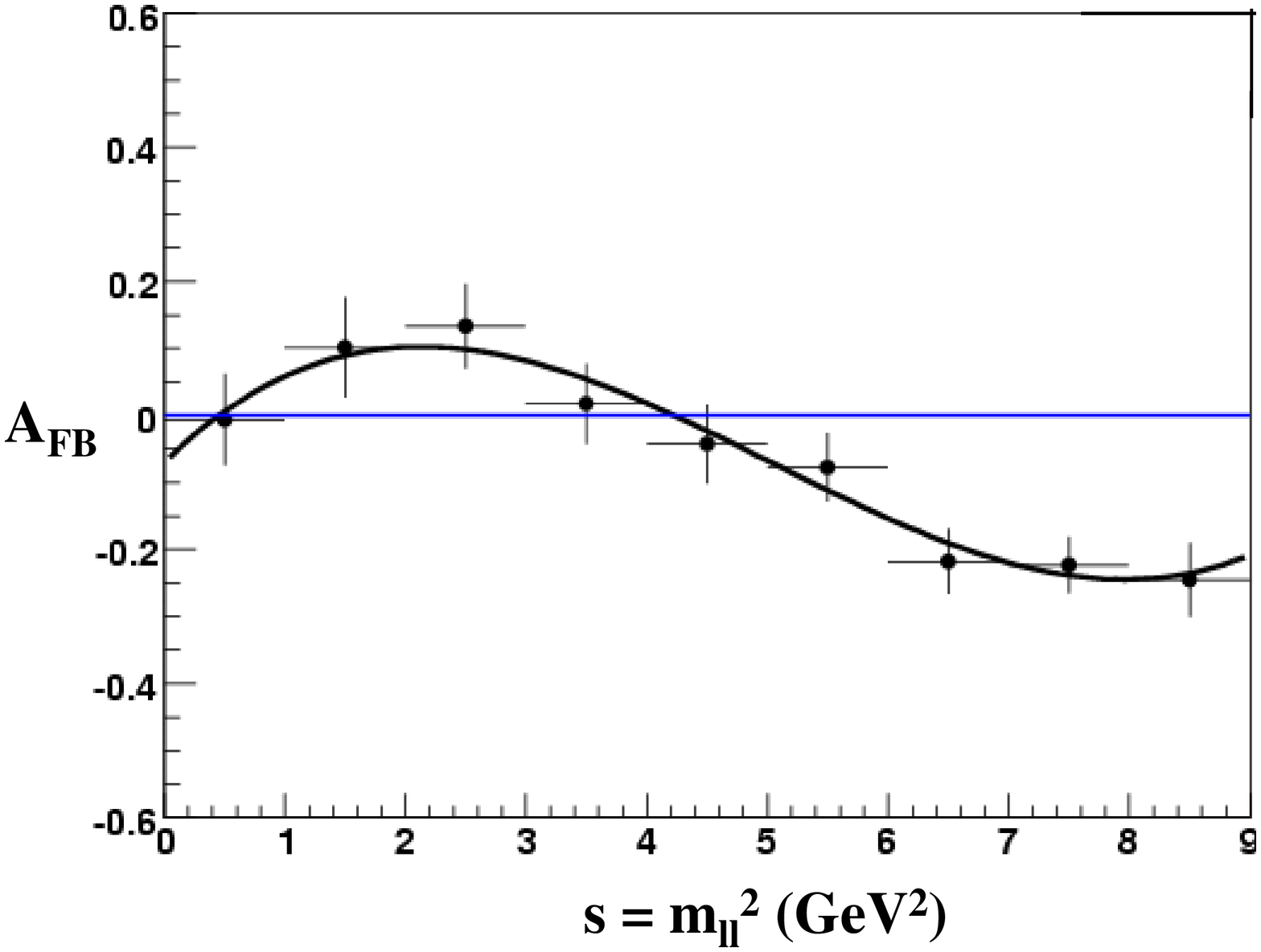,scale=0.35}
\psfig{figure=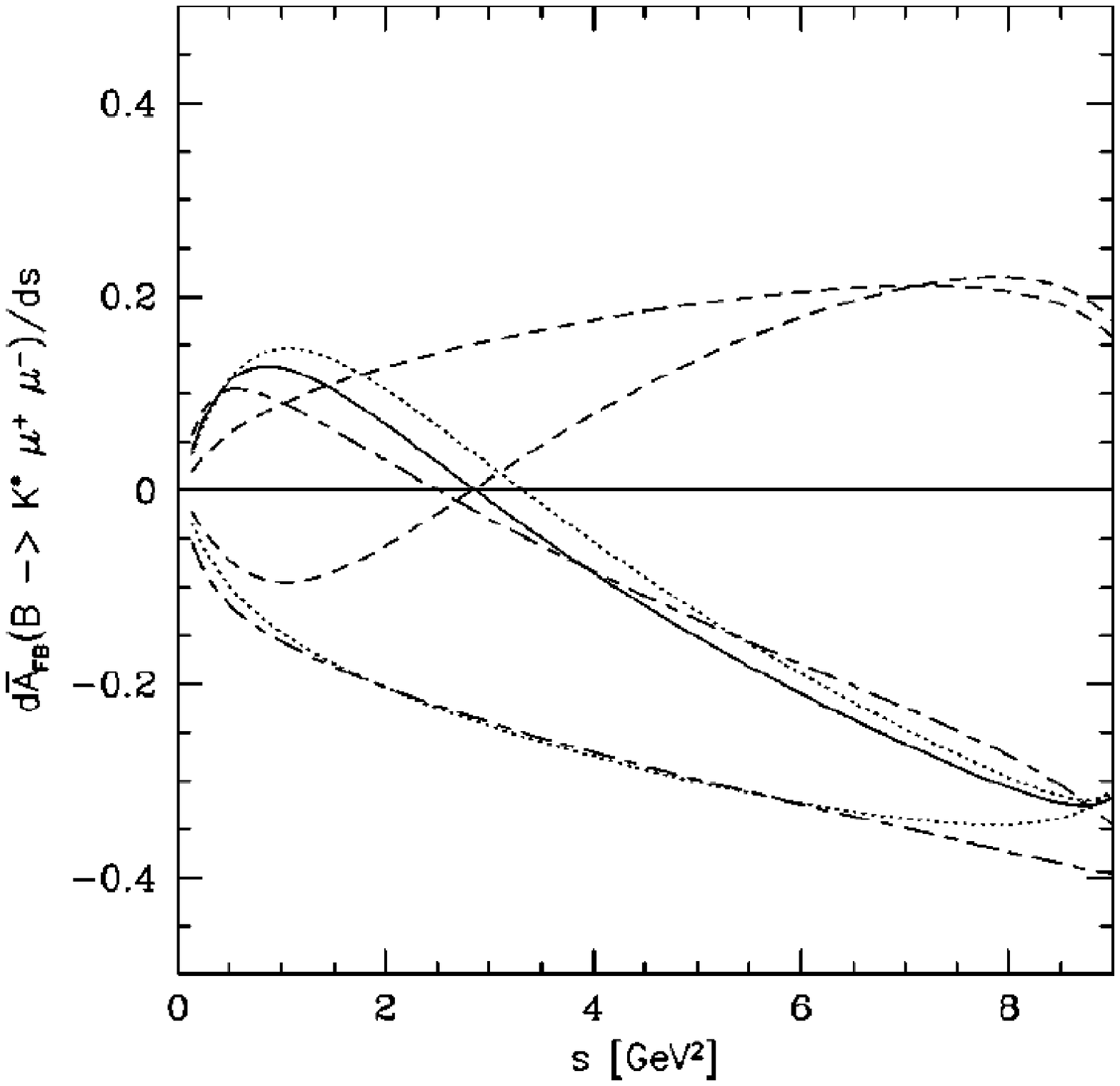,scale=0.35}
 \caption{Dimuon forward-backward asymmetry as a function of the squared dimuon invariant mass. The left plot shows the experimental result while the right plot shows theoretical predictions for several working hypotheses. Right plot: the solid line is for the SM-based prediction and the dotted and dashed lines represent results from different SUSY-based scenarios where the signs of $C_{10}$ and $C_7^{eff}$ Wilson coefficients are flipped.\label{fig:AFB_kstarmumu}}
\end{figure}

This transition is governed mostly by electroweak and
electromagnetic penguin operators, $O_7$, $O_9$ and $O_{10}$. The
rate is dominated by $|C_9^{eff}|^2$, $|C_{10}|^2$ and the sign of
$C_7^{eff}$. An interesting observable is the leptons
forward-backward asymmetry which is highly sensitive to the
relative sign of $C_7^{eff}$ and $C_{10}$. In the leptons pair
rest frame, we consider the angle $\theta_{ll}$ of the leptons
with respect to the $B$ meson momentum. The asymmetry is then
defined as:
\begin{equation}
A_{FB}(\hat{s}=\frac{m^2_{ll}}{m^2_b})=\frac{\int_0^1
dcos\theta_{ll}\frac{d^2\Gamma(B\to X_s
l^+l^-)}{dcos\theta_{ll}d\hat{s}}-\int_{-1}^0dcos\theta_{ll}\frac{d^2\Gamma(B\to
X_s l^+l^-)}{dcos\theta_{ll}d\hat{s}}}{\int_0^1
dcos\theta_{ll}\frac{d^2\Gamma(B\to X_s
l^+l^-)}{dcos\theta_{ll}d\hat{s}}+\int_{-1}^0dcos\theta_{ll}
\frac{d^2\Gamma(B\to X_s l^+l^-)}{dcos\theta_{ll}d\hat{s}}}
\end{equation}
The point $\hat{s}_0$ where this quantity cancels to zero has a
particular theoretical interest since it is known with a
reasonable accuracy \cite{AliKll}.\\
Studies have been performed for the decay $B_d\to K^{*0}
\mu^+\mu^-$ \cite{Kstarmumu_Lhcb}. Particular care has been taken
to apply selection cuts that don't bias the dimuon invariant mass
distribution. 7.2k signal events are expected for an integrated
luminosity of $2~fb^{-1}$. Figure \ref{fig:AFB_kstarmumu} shows
the resulting expected asymmetry distribution, along with
theoretical predictions on the shape (taken from reference
\cite{AliKstll}).
The remarkable feature of the predictions lies in the fact that,
beside the result that the zero point of the asymmetry is known to
a precision of $\sim0.6~GeV^2$ in the Standard Model, models where
$C_7^{eff}>0$ predict that the asymmetry does not cancel. In the
same studies, other variables with minimal theoretical
uncertainties, such as the fraction of longitudinal polarization
of the $K^{*0}$, have been considered as interesting probes and
are expected to provide further sensitivity to New
Physics.\\
The analysis of $B^+\to K^+ l^+l^-$ modes (with $l=e,~\mu$) has
also been considered \cite{Kplusll_Lhcb}. It has been shown
\cite{Hiller_Kruger} that the following ratio:
\begin{equation}
R_X = \frac{\int_{s_{min}}^{s_{max}} ds\frac{d\Gamma(B\to X
\mu^+\mu^-)}{ds}}{\int_{s_{min}}^{s_{max}} ds\frac{d\Gamma(B\to X
e^+e^-)}{ds}}
\end{equation}
, can be predicted with a very good precision in the Standard
Model. In particular for $X=K$, this ratio is equal to one at the
$10^{-4}$ level. Substantial deviations from this value could
occur from scalar $\sim\bar{s}Rb\bar{l}l$ and pseudo-scalar
$\sim\bar{s}Rb\bar{l}\gamma^5l$ operators contributing to the
effective hamiltonian. The corresponding Wilson coefficients
include the lepton masses and are therefore responsible for a
possible difference between electron and muon modes.\\
Experimentally, selections have been optimized to reject
backgrounds of type $Xl^+l^-$ with badly reconstructed $X$ and it
has been shown that the sensitivity to $R_K$ reaches few percent
with an integrated luminosity of $10~fb^{-1}$.
\section{$B_s\to\mu^+\mu^-$}
\begin{figure}[!h]
\centering \psfig{figure=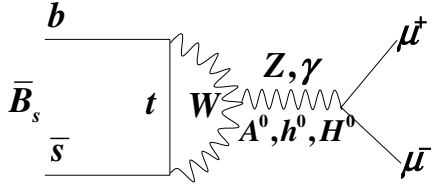,scale=0.25}
 \caption{Annihilation diagram of the decay $B_s\to\mu^+\mu^-$. In MSSM or
Double Higgs Model, Z boson or $\gamma$ mediation can be replaced
by neutral Higgs bosons, as depicted.\label{fig:bstomumu_diag}}
\end{figure}

This rare mode is mediated by second order annihilation diagrams
such as the one shown in figure \ref{fig:bstomumu_diag}.

\begin{figure}
\centering \psfig{figure=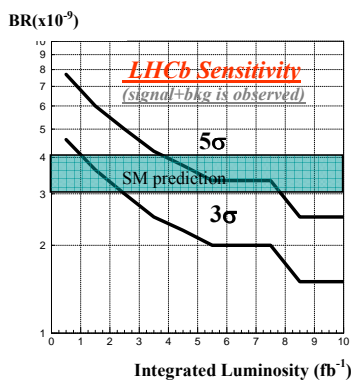,scale=0.8}
\caption{Observation ($3\sigma$) and Discovery ($5\sigma$) limits
for $B_s\to\mu^+\mu^-$ as a function of integrated luminosity in
case where both signal and background are observed.
\label{fig:bstomumu_sens}}
\end{figure}

It is suppressed by a factor $\sim m_{\mu}^2/m_B^2$ in the
Standard Model, leading to a branching ratio of $\sim
3\times10^{-9}$, \cite{bstomumu_theo}. A possible enhancement
could occur through neutral Higgs mediation in constrained Minimal
Super Symmetry or Two Higgs Models with large $tan\beta$
\cite{bstomumu_theo}. In that case, phenomenology predicts
$\Gamma(B_s\to\mu^+\mu^-)\propto\frac{m_b^2m_{\mu}^2tan^6\beta}{M_{A^0}^4}$.\\
On the experimental side \cite{bstomumu_exp}, the decay is easy to
reconstruct but is embedded in a huge background coming from
leptonic $b$ decays. Sensitivity studies have been performed to
test the discovery power as a function of statistics. Figure
\ref{fig:bstomumu_sens} shows the results. Early observation with
$2~fb^{-1}$ is possible for SM-like rates while discovery can be
envisaged with even lower statistics if New Physics enhances the
branching ratio.
\section*{Conclusion}
The key studies reviewed in this paper reflect the New Physics
sensitivity timeline for the LHCb experiment. First data
(integrated luminosity $<\sim0.5 fb^{-1}$) will give us first
answers on any substantial enhancement of Standard Model
suppressed observables, such as the weak mixing phase $\beta_s$ in
the $B_s$ oscillations or the rate of $B_s\to\mu^+\mu^-$. The
$2~fb^{-1}$ milestone will then consolidate the first
observations. Final data sample of what one could qualify as a
first "phase", $\sim 10 fb^{-1}$, is expected to give us more
insight on the flavour structure of New Physics through precise
measurements of rates and $CP$ asymmetries and will also allow
more significant determination of differential rate asymmetries
and polarizations in rare processes.

\end{document}